\documentclass[twocolumn,floatfix,showpacs,pra,aps,letterpaper]{revtex4}
\usepackage{graphicx}
\usepackage{multirow}

\begin{document}

\newcommand{\CNOT}{\textsc{cnot}}

\title{Optimal correction of concatenated fault-tolerant quantum codes}
\author{Z. W. E. Evans$^{1}$}
\author{A. M. Stephens$^{1,2}{}\footnote{electronic address: astephens@nii.ac.jp}$}
\affiliation{$^{1}$School of Physics, The University of Melbourne, Victoria 3010, Australia}
\affiliation{$^{2}$National Institute of Informatics, 2-1-2 Hitotsubashi, Chiyoda-ku 101-8430, Japan}

\date{\today}

\begin{abstract}
We present a method of concatenated quantum error correction in which improved classical processing is used with existing quantum codes and fault-tolerant circuits to more reliably correct errors. Rather than correcting each level of a concatenated code independently, our method uses information about the likelihood of errors having occurred at lower levels to maximize the probability of correctly interpreting error syndromes. Results of simulations of our method applied to the [[4,1,2]] subsystem code indicate that it can correct a number of discrete errors up to half of the distance of the concatenated code, which is optimal.
\end{abstract}

\pacs{03.67.Lx, 03.67.Pp}
\maketitle

\section{Introduction}
The expectation that quantum computers will, one day, outperform classical computers is founded on quantum error correction, which will be necessary to mitigate environmental decoherence and errors that arise from imprecise quantum control. Various methods for error correction are now known---for example, topological error correction~\cite{Kitaev1,Raussendorf2}---but the first and most prominent method involves concatenating \cite{Knill4} small quantum codes such as the Shor code~\cite{Shor4} and the Steane code~\cite{Steane1}.

Quantum codes work by redundantly encoding the state of a single physical qubit in the combined state of a few qubits, whereafter errors in the states of some subset of these qubits can be diagnosed and corrected using a circuit that is designed to obtain the error syndrome whilst limiting the propagation of errors~\cite{Nielsen1}. However, because this circuit is inevitably formed from a number of error-prone physical locations, such as quantum gates and quantum measurements, there are more places from which errors can arise than before. This tradeoff implies that the error rate of an encoded qubit will only be lower than the error rate of a physical qubit if the error rate of the physical locations is below a certain threshold~\cite{Aharonov1,Aharonov2,Knill5,Preskill1,Aliferis1}.

A single level of encoding and error correction can only reduce the error rate by so much. A greater reduction can be achieved by encoding a single physical qubit in the state of a few qubits, as before, and then encoding each of those qubits in the state of more qubits, and so on. This forms a concatenated quantum code, which allows, in theory, an arbitrarily long and large universal quantum computation to be undertaken reliably. However, as the number of qubits required increases exponentially with the number of concatenation levels, it is likely that only a few levels will be practical. It is important, therefore, to make each level of error correction as effective as possible.

Typically, each level of a concatenated code is corrected independently of all other levels with the implicit assumption that errors are equally likely on all qubits. For $l$ levels of a distance-$d$ code, this method results in a concatenated code that can fail if there are $\lceil d/2\rceil^l$ errors. However, as the total distance of the concatenated code is $d^l$, one might expect that a more sophisticated correction method may exist, one that will never fail with fewer than $\lceil d^l/2\rceil$ errors. Such a method is known in the case where the error syndrome can be obtained perfectly~\cite{Poulin1} but not in the more general case where the error syndrome is unreliable. 

Following our previous work improving the correction method for two levels of a specific code~\cite{Evans1}, here we present a more general correction method that is designed to reliably correct up to $\lceil d^l/2\rceil$ errors. Our method works with existing CSS stabilizer codes~\cite{Calderbank2,Steane2} and with existing fault-tolerant circuits, and takes advantage of the fact that the process of error correcting an encoded qubit yields information that can be used to estimate the probability of that encoded qubit having an error. For example, whenever we observe a non-zero syndrome we know that at least one error has occurred, and so our confidence in the relevant encoded qubit should be decreased. Here we use this information to determine the most likely sets of errors that could give rise to the syndromes we observe at each level of encoding. 

We describe our method in general and also applied to the [[4,1,2]] subsystem code~\cite{Bacon1}. To test our method we perform Monte Carlo simulations of up to four levels of error correction and find that it performs as expected. Our method has also been used in a study of the threshold for fault-tolerant quantum computation on a linear array of qubits with only nearest-neighbor interactions~~\cite{Stephens1}. It remains to be proven that our method works for an arbitrary number of levels of error correction.

\section{Correction method}

It is useful to begin by outlining the general structure of a quantum circuit that is protected by a concatenated fault-tolerant quantum code. First, all of the qubits in the circuit are replaced by encoded qubits. Then, all of the \emph{locations} in the circuit---such as quantum gates, quantum measurements, and quantum memory---are replaced by encoded locations, which perform the corresponding operations on encoded qubits. Interspersed between encoded locations are \emph{error-correction cycles}, which involve performing a fault-tolerant circuit to obtain an error syndrome and then using a classical algorithm to determine what, if any, correction should be applied to the relevant encoded qubit. This correction is applied to one or more of the physical qubits that make up the encoded qubit. As described, the original circuit is now an encoded circuit protected by one \emph{level} of error correction. Concatenation involves repeating this procedure, so that all of the physical locations---that is, all of the physical locations that make up the encoded locations and the error-correction circuits---are themselves replaced by encoded locations and error-correction cycles. This introduces a second level of error correction and now corrections may be applied to encoded qubits as well as physical qubits.

In this context, our correction method replaces the classical algorithm that is used in each error correction cycle. Our algorithm is general, meaning it is the same for each error correction cycle at each level. We assume a discrete error model whereby the probability of error at each physical location, $p$, is constant and independent of the other locations. We also assume that the quantum computer will be operating in the low-$p$ limit, thus the value of the smallest exponent of $p$ is sufficient to describe probabilities. It may be possible to improve the method by relaxing these assumptions.

\subsection{Flags}
For each location in the syndrome circuit we determine what effect a discrete $X$ or $Z$ error occurring at that location would have at the end of the circuit. The error may affect one or more of the data qubits that make up the encoded qubit, one or more of the ancilla qubits that are measured to obtain the syndrome, or some combination of these. Taking into account that some errors are equivalent to other errors, we assign a \emph{flag} to each unique error. In other words, there is a flag for each effect that can be caused by a single error. A flag represents the possibility of a single error. The probability that a flag represents an actual error is described by the \emph{weight} of the flag such that $\textrm{Pr(error)} = \mathcal{O}(p^{\textrm{weight}})$. Since we assume that errors occur independently at each location, the probability of a set of flags representing a set of actual errors is $\mathcal{O}(p^{\sum{\textrm{weight}}})$, where the sum is over the set of flags. Errors and flags in the $X$ basis are separate from those in the $Z$ basis.

An error on a data qubit will eventually have its effect transferred to one or more ancilla qubits. The location of the error determines which ancilla qubits will be affected. But, an error that affects a data qubit after it has interacted with the ancilla qubits cannot be detected until the next error-correction cycle. Flags that represent such errors are called \emph{transitive flags}. Transitive flags are put aside until the current error-correction cycle is complete, after which each transitive flag is converted to its corresponding non-transitive flag for use in the next error-correction cycle.

Initially, all flags can be thought of as having infinite weight, as there cannot be any errors before any operations have been done. When a circuit location is executed, the weight of the flag corresponding to an error at that location is updated to be the minimum of its current value and a \emph{confidence rating} reported by that location. At the physical level, the confidence rating reported by each location is one, reflecting our assumption that physical locations fail with probability $p$. The confidence rating reported by each encoded location is determined during the error-correction cycle that follows that location, based on the syndrome and the weights of the flags associated with that location (see $\S$2.2). Detecting errors at the physical level will lower the confidence rating reported by locations all the way up to the highest level of encoding.

Two-qubit locations are special in that errors may be shared by both qubits. For example, an error occurring before a $\CNOT$ may be copied and an error occurring during a $\CNOT$ may affect both the control and the target qubits. However, each encoded qubit has its own set of flag weights representing the possibility of errors on the data qubits that make up that encoded qubit. Therefore, during a two-qubit location that copies errors, flags belonging to the encoded qubit from which errors may be copied are used to update the corresponding flags belonging to the encoded qubit to which errors may be copied. Also, confidence ratings are used to update flags that correspond to correlated errors (or, equivalently, errors that are copied by two-qubit locations). Specifically, the weight of the flag corresponding to the correlated error associated with a two-qubit location is updated to be the maximum of the confidence ratings reported by the two error-correction cycles following that location. The maximum of the two confidence ratings is appropriate since a correlated error involves errors affecting both qubits. The sum of the two confidence ratings would correspond to two uncorrelated errors. Again, the weight of each flag is updated only if its value would be lowered from its current value. The flags corresponding to uncorrelated errors are updated as usual.

\subsection{Flag matching, error correction, and confidence reporting}
During each error-correction cycle the ancilla qubits are measured to obtain a syndrome. Each possible syndrome is consistent with a finite number of sets of errors and each set of errors has a corresponding set of flags. Each set of flags that matches the syndrome is called a \emph{flag match}. The most likely cause of the syndrome is the set of errors that is represented by the flag match with the lowest weight. With this flag match identified, we apply corrections to whichever data qubits would be affected by this set of errors.

Different flag matches will imply different corrections. In particular, there will always be a match which implies a complementary set of corrections to that of the match with the lowest weight---that is, the corrections of both matches combine to form an encoded operator. If we correct based on one flag match but the true set of errors is represented by its \emph{complement match} then the result is an encoded error. To determine the confidence of the encoded location we consider the probability that our choice of correction will result in an encoded error,
\begin{equation}
  \textrm{Pr(fail}\vert\textrm{syndrome}\cap\textrm{flags})=\frac{\textrm{Pr(fail}\cap\textrm{syndrome}\vert\textrm{flags})}{\textrm{Pr(syndrome}\vert\textrm{flags})}.
	\label{eq: conditional prob}
\end{equation}
Pr(fail$\vert$syndrome$\cap$flags) is the probability of an encoded error given the current syndrome and set of flags. Pr(syndrome$\vert$flags) is the probability of the current syndrome occurring given the current set of flags, regardless of success or failure. Pr(fail$\cap$syndrome$\vert$flags) is the probability of the current syndrome occurring and error correction resulting in an encoded error. In the low-$p$ limit these probabilities are dominated by the leading-order terms and the coefficients are not important. Since our method applies corrections based on the most likely cause of the syndrome, to leading order Pr(syndrome)$=$Pr(success$\cap$syndrome). This implies
\begin{equation}
  \frac{\textrm{Pr(fail}\cap\textrm{syndrome}\vert\textrm{flags})}{\textrm{Pr(success}\cap\textrm{syndrome}\vert\textrm{flags})} \approx p^{y-x},
	\label{eq: failure first order}
\end{equation}
where $x$ is the weight of the flag match with the lowest weight and $y$ is the weight of its complement match. Therefore, 
\begin{equation}
  C_{EN} = y - x
	\label{eq: logical confidence}
\end{equation}
is the confidence rating determined during this error-correction cycle. The confidence rating represents the probability of an encoded error, $\mathcal{O}(p^{C_{EN}})$. 

There is also a possibility that errors, and consequent corrections, will result in a state that is outside of the code space, so that it is neither the correct state nor the state affected by an encoded error. Just as we use the complement match to determine the probability of an encoded error, we can determine the probability of errors on the individual data qubits by considering other flag matches. For large codes there will also be probabilities of sets of correlated data errors that we can determine. Let $E$ represent a particular set of errors after an error-correction cycle. Then
\begin{equation}
  \textrm{Pr(}E\vert\textrm{syndrome}\cap\textrm{flags}) = \frac{\textrm{Pr(}E\cap\textrm{syndrome}\vert\textrm{flags})}{\textrm{Pr(syndrome}\vert\textrm{flags})}.
	\label{eq: correction flagging}
\end{equation}
The weight of the flag corresponding to this set of errors should, therefore, be updated to be the minimum of its current weight and the weight of each set of flags that match this outcome, taking into account any corrections that have been applied, minus the weight of the minimum-weight flag match. (See $\S$3.1 for an example.) 

Through the careful consideration of how errors propagate in error-correction circuits, we can identify every possible cause of every syndrome and thus accurately calculate confidence ratings for every qubit at every level of encoding. This allows us to always correct based on the most likely set of errors and makes each level of error correction as effective as possible at reducing the encoded error rate.

\section{Application to the [[4,1,2]] subsystem code}
\label{sec: application}
Here we apply our correction method to the [[4,1,2]] subsystem code~\cite{Bacon1}. The stabilizer group of this code is generated by the operators
\begin{equation}
\begin{array}{c}
X_1X_2X_3X_4,\\
Z_1Z_2Z_3Z_4,
\end{array} 
\label{eq: stabilizers}
\end{equation}
where $X_i$ and $Z_i$ respectively represent the Pauli operators $\sigma_X$ and $\sigma_Z$ applied to the $i^{th}$ qubit. Tensor products and identity operators are implicitly present. The gauge group of this code is generated by the operators
\begin{equation}
\begin{array}{c}
 X_1 X_2,\\
 X_3 X_4,\\
 Z_1 Z_3,\\
 Z_2 Z_4.
\end{array} 
\label{eq: gauge}
\end{equation}
The encoded $X$ and $Z$ operators are
\begin{equation}
\begin{array}{c}
X_L = X_1 X_3,\\
Z_L = Z_1 Z_2,
\end{array} 
\label{eq: logical ops}
\end{equation}
or equivalently any operators that are formed by combining (\ref{eq: logical ops}) and (\ref{eq: gauge}).

The syndrome circuit that we have chosen to use, along with the flags that will be used in our correction method, is shown in Fig.~\ref{fig: error map}~\cite{Aliferis4}. The circuit measures the gauge operators, two at a time. Since the stabilizers are products of gauge operators, the combined parity of each pair of measurement results is effectively the result of measuring of a stabilizer. Individually, each measurement tells us nothing about the syndrome.

\subsection{Correction method}
\begin{figure*}
\begin{center}
\resizebox{90mm}{!}{\includegraphics[angle=90]{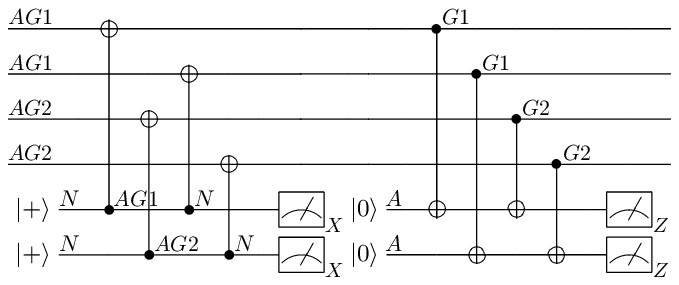}}
\end{center}
\vspace*{-17pt}
\caption{Syndrome circuit for the [[4,1,2]] code, with $X$ flag mapping shown. The error syndromes in the $X$ and $Z$ bases are obtained by performing operator measurements of the gauge operators in (\ref{eq: gauge}). The parity of the two $X$ gauge operator measurements gives the $Z$ error syndrome, and vice versa. The letters on the circuit mark out regions associated with each of the $X$ flags. The confidence rating of each circuit location is associated with the flag that is labeled to the left of that location. (See Table 1 for the flag labels). For example, an $X$ error that occurs in the region marked $AG1$ will affect one of the qubits involved in the first $X$ gauge operator and the effect will be seen on the ancilla used to perform the operator measurement. An $X$ error that occurs in the region marked $A$ will flip the parity of the ancilla but will not affect the data at all. An $X$ error that occurs in the $N$ region will have no effect on the data or the observed syndrome. The flag regions shown here are for $X$ flags only but the $Z$ flag map is similar.}
\label{fig: error map}
\end{figure*}

The syndrome of the code consists of a single bit for each basis. A syndrome of zero indicates that no error was detected; one indicates that an error was detected but provides no information about which qubits were affected by the error. When the syndrome of one is measured, in the absence of any additional information we must simply guess which data qubit was affected by the error. Each data qubit has a partner qubit in each basis for which errors are equivalent. For example, an $X$ error on the first qubit is the same as an $X$ error on the second qubit up to the application of the $X_1 X_2$ gauge operator. Such an error can, therefore, be corrected by the application of an $X$ operator on either the first or second qubit. To simplify the analysis of errors, we refer to errors by which gauge operator they are a part of, rather than by which data qubit they affect. An $X$ error on the first or second qubit is said to be an error on the \emph{first gauge} in the $X$ basis, a $Z$ error on the second or fourth qubit is said to be an error on the \emph{second gauge} in the $Z$ basis, and so on. When there is an error on one gauge and the syndrome is misinterpreted so that the correction operation is applied to the other gauge the result is always an encoded error.

In order to reliably correct errors using the [[4,1,2]] code, extra information is required to distinguish between errors affecting the first and second gauges. One method for obtaining this information is to pass classical messages describing the relative probabilities of errors from lower levels of error correction to higher levels. Previously, it was shown that a scheme such as this could increase the minimum number of errors that cause failure to a number that scales according to the Fibonacci series with the number of levels of error correction, $l$~\cite{Knill1,Aliferis2}. In contrast, the optimal scaling is given by $2^{l-1}$.

For the [[4,1,2]] code, using the circuits we have chosen, there are five unique effects that can result from a single error. Therefore there are five flags that we are required to track for each encoded qubit at each level of encoding. These flags are defined in Table 1. Note that two of the flags are transitive. The set of error locations belonging to each flag are shown in Fig.~\ref{fig: error map}. 

\begin{table}
\begin{center}
\begin{tabular}{c|c}
  flag & effect\\
  \hline
  $AG1$ & first gauge and ancilla\\
  $AG2$ & second gauge and ancilla\\
  $A$ & ancilla only \\
  $G1$ & first gauge only (transitive)\\
  $G2$ & second gauge only (transitive)
\end{tabular}
  \label{tab: flags}
\caption{Names assigned to the flags of the [[4,1,2]] code. There are five flags for each basis. \emph{First gauge} refers to an error on either of the two qubits involved in the first gauge operator of the appropriate basis in (\ref{eq: gauge}), similarly for \emph{second gauge}. \emph{Ancilla} refers to an error which flips the parity of the two gauge operator measurements---that is, an error which changes the outcome of the syndrome measurement. Flags $G1$ and $G2$ become $AG1$ and $AG2$ respectively at the end of the error-correction cycle as they will be seen on the ancilla of the following error-correction cycle.}
\end{center}
\end{table}

The errors represented by the flag match with the lowest weight are identified as being the most likely cause of the observed syndrome. Corrections are applied to any data qubits marked by the flag match, then flags are updated accordingly. Table 2 shows a list of all possible flag matches along with the various confidence ratings that would result from each of them being acted on.

\begin{table*}
\begin{center}
\begin{tabular}{c|c||c|c|c|c}
  synd. & flag match & corr. & $C_{G1}$ & $C_{G2}$ & $C_{EN}$ \\
  \hline
 \multirow{4}{*}{0}
   & no flags & none & $AG1+A$ & $AG2+A$ & $AG1+AG2$\\
   & $AG1+A$ & \multicolumn{4}{c}{N/A}\\
   & $AG2+A$ & \multicolumn{4}{c}{N/A}\\
   & $AG1+AG2$ & \multicolumn{4}{c}{N/A}\\
   \hline
 \multirow{4}{*}{1}
  & $AG1$ & $G1$ & $A-AG1$ & $AG2+A$ & $AG2-AG1$\\
  & $AG2$ & $G2$ & $AG1+A$ & $A-AG2$ & $AG1-AG2$\\
  & $A$ & none & $AG1-A$ & $AG2-A$ & $AG1+AG2$\\
  & $AG1+AG2+A$ & \multicolumn{4}{c}{N/A}
  \label{tab: flag matches}
\end{tabular}
\caption{For each of the two syndromes (column 1) there are four possible flag matches (column 2). Corrections (column 3) and flag updates (columns 4-6) depend on which flag match has the lowest weight. The flag matches marked with N/A always have a weight equal to or higher than other matches and so will never be acted on. $C_{G1}$ and $C_{G2}$ are confidence ratings for the transitive flags $G1$ and $G2$ at the current level of error correction. $C_{EN}$ is the confidence rating for the next-highest level of error correction. In the expressions for evaluating the weights of the flags matches (column 2) and for calculating the confidence ratings (columns 4-6) the symbols $A$, $AG1$, and $AG2$ represent the weights of those flags.}
\end{center}
\end{table*}

As an example, let us assume that the $X$ basis syndrome is measured to be one. This could be because of an error affecting the first gauge, an error affecting the second gauge, an error affecting the ancilla, or the combination of all of the above. These cases are represented by rows 5-8 of Table 2. First, we determine which of these is most likely by considering the weights of the relevant flags, $AG1$, $AG2$, and $A$. Let us assume that the weight of $AG2$ is lower than the weights of both $AG1$ and $A$ and thus an error affecting the second gauge is most likely. This case is described by the sixth row of Table 2. According to the third column, as the flag match indicates that an error has affected the second gauge, a correction should be applied to the second gauge---that is, an $X$ correction should be applied to either qubit three or qubit four. Then, the confidence ratings for the transitive flags should be calculated. These transitive flags represent the probabilities of errors on the outgoing data qubits and will be used in the next error-correction cycle. Since the observed syndrome is one and we have applied a correction to the second gauge, for \emph{only} the outgoing first gauge to be affected by an error there would have to have originally been errors affecting both gauges and the ancilla. The probability of this is described by the weight $AG1+AG2+A$. According to the fourth column, the confidence rating for the first gauge is equal to the conditional probability of this occurring, or the weight $AG1+AG2+A-AG2=AG1+A$. For \emph{only} the outgoing second gauge to be affected by an error there would have to have originally been an error affecting the ancilla. The probability of this is described by the weight $A$. This means there was no error affecting the second gauge in the first place, but one has been added by our correction. According to the fifth column, the confidence rating for the second gauge is the weight $A-AG2$. Finally, we need to determine the conditional probability of an encoded error, which is an error affecting \emph{both} gauges. Since our correction is applied to the second gauge, the complement match is an error affecting the first gauge. The probability of this is described by the weight $AG1$. According to column six, the confidence rating for the encoded qubit is the weight $AG1-AG2$. This confidence is then reported by the preceding encoded location and used in the next-highest level of error correction.

\subsection{Simulation and results}
To test our method we simulate a $\CNOT$ extended rectangle~\cite{Aliferis1} including all physical locations. The correction method is simulated alongside the error-correction circuits so that it operates in the same way as it would in a real quantum computer. The $\CNOT$ is defined to succeed only if the measurement results of both logical qubits in both bases would be correct. The simulations use the method outlined in~\cite{Evans1}, which involves keeping track of discrete errors as they propagate through the circuit. We assume a stochastic error model, where memory, initialization, readout, single-qubit gate, and two-qubit gate errors are all equally likely to occur. Errors are randomly selected one- and two-qubit discrete Pauli errors.

\begin{figure*}
\begin{center}
\resizebox{95mm}{!}{\includegraphics*{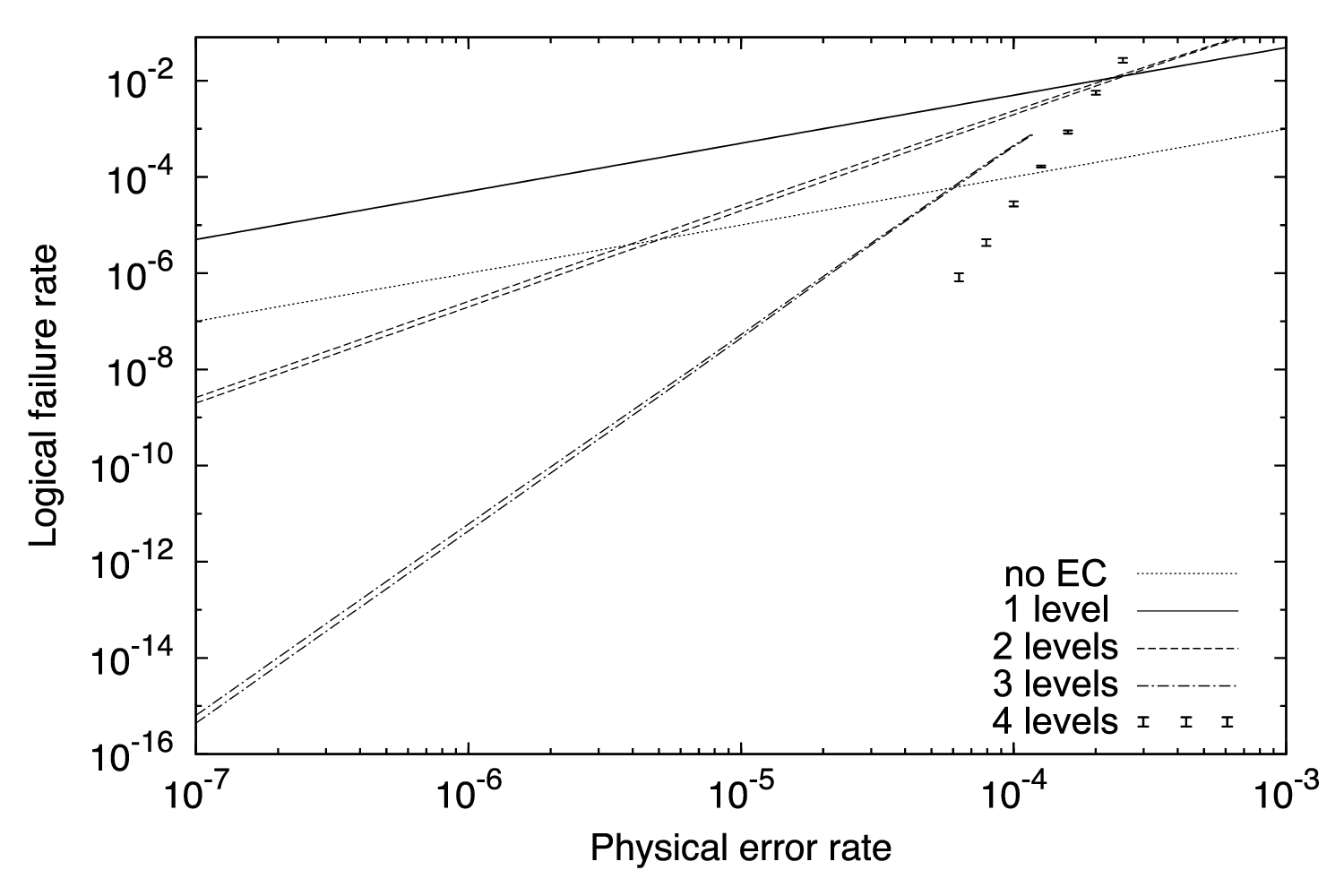}}
\end{center}
\vspace*{-20pt}
\caption{Failure rate of $\CNOT$ extended rectangle as a function of physical error rate for one, two, three, and four levels of error correction with the [[4,1,2]] code using our correction method. $2\sigma$ of statistical error above and below the mean is shown for each set of data. As expected, the minimum number of discrete errors that can cause failure is one, two, and four for one, two, and three levels of error correction respectively. This can be inferred by the gradients of the lines in the low-$p$ region. For four levels of error correction, the gradient of the series in the region around $p=10^{-4}$ cannot be used to estimate the lowest order of Eq.~(\ref{eq: error expansion}) and hence the minimum number of errors that can cause failure. This is because our data at this physical error rate is dominated by combinations of 16 or more errors. In our data for four levels of error correction there were no cases that could have failed due to seven or fewer errors, as expected. Note that the asymptotic threshold appears to be around $p=2\times10^{-4}$. Also, since the first level of the code is unable to reliably correct any number of errors, the logical failure rate with one level of error correction is always greater than the physical error rate.}
\label{fig: results}
\end{figure*}

We are primarily interested in the logical failure rate when the physical error rate, $p$, is low. Directly simulating the circuit in the low-$p$ limit requires too many runs to obtain statistically significant results. Instead, for up to three levels of error correction, we use the expansion of logical failure rate with respect to the physical error rate,
\begin{equation}
  p_{L} = \sum^N_{i=0}{r_i {N\choose i}p^i(1-p)^{N-i}},
  \label{eq: error expansion}
\end{equation}
where $N$ is the number of physical locations in the error-correction circuit, $p$ is the physical error rate, and $r_i$ is the probability of logical failure after error correction given exactly $i$ errors. To estimate values of $r_i$ we simulate the circuit with exactly $i$ errors placed randomly, repeating many times for each $i$. We approximate $p_L$ by truncating the series above $i = 29$. This approximation breaks down for values of $p$ for which the probability of having more than 29 errors becomes significant. For four levels of error correction we can only obtain statistically significant results for relatively high values of $p$. To do this we simulate the circuit directly. The failure rate of the logical $\CNOT$ as a function of physical error rate is plotted in Fig.~\ref{fig: results} for one, two, three, and four levels of error-correction.

\section{Further work}
As described, our method assumes that each physical location is equally likely to fail. However, this need not be the case. Relative probabilities of failure for individual physical locations are naturally represented in by the confidences at the physical level. Rather than setting the confidence of all physical locations to one, we can choose to assign different confidence ratings to each type of operation or each individual physical device. These confidence ratings could be based on data obtained by characterization of the gates. Alternatively, data collected during error correction could be fed back to dynamically update confidence ratings of physical locations in a similar way to how our method updates confidence ratings of encoded locations.

Our method could be applied to a post-selection based scheme. There are several ways this could be done, since we are free to choose the confidence threshold which determines whether ancillary states are accepted or rejected. To only accept states that have the maximum confidence would reduce the scheme to ordinary post-selection, as a state would be rejected if any errors are detected at any level regardless of whether or not these errors are correctable at higher levels. Accepting all states that have a confidence greater than one would result in a high acceptance rate, but would not be optimal with respect to logical fidelity. In any case, the confidence of the states that are accepted could be used in higher-level error correction following post-selection. There may also be some advantage to using the confidence ratings determined at the highest level of error correction in some quantum algorithms.

Finally, our method can be applied to higher distance subsystem codes and to other CSS stabilizer codes. It would be interesting to investigate the pros and cons of using different codes with our method and also to rigorously establish the specific requirements on codes and circuits that enable us to achieve optimality with respect to the minimum number of errors that cause the code to fail.

\section*{Acknolwedgements} Our simulator uses the SIMD-oriented Mersenne Twister pseudorandom number generator~\cite{Saito1}. We acknowledge financial support from the Australian Research Council, the US National Security Agency, and the Army Research Office under contract number W911NF-04-1-0290.

\bibliographystyle{unsrt}

\end{document}